\documentstyle[12pt,epsfig]{article}
\begin{document}
\begin{center}
\vspace*{10pt}
{\Large \bf Recent and Future Measurements} \\
\vspace*{10pt}
{\Large \bf  of the Neutron Electric Form Factor} \\
\vspace*{20pt}
{\large Andrei Semenov}\\
\vspace*{0.10cm}
  {\it University of Regina, Saskatchewan, S4S~0A2 Canada} \\
\vspace*{0.10cm}
\vspace*{10pt}
\end{center}

\begin{center}
\vspace*{0.40cm}
I review recently
conducted measurements of $G_{En}$ as well as precision form\\ 
factor experiments at high momentum transfer that will be performed with\\
11 GeV electron beam at Jefferson Lab.
\end{center}

\section{Introduction}
Measurements of the electromagnetic nucleon form factors are of great importance in
trying to understand the quark substructure of the nucleon. They provide important
constraints on generalized quark distributions (GPDs) that compliment
one-dimensional picture of hadron structure as a distribution of longitudinal
momentum and helicity of partons with spatial information in the plane perpendicular
to the direction of movement ("nucleon tomography").  GPDs fits to the Pauli form
factors of proton and neutron allow to estimate helicity flip distributions, which
play an essential role in understanding how the total spin of the nucleon is made up
from quarks and gluons. Since proton form factors, for which data are abundant, are
dominated by $u$ quarks, new high-quality data on the neutron form factors in a wide
$Q^2$-range (especially, for electric form factor of neutron, $G_{En}$ above 3
(GeV/c)$^2$) would be highly valuable for pinning down the expected drastic
differences in the spatial distribution of $u$ and $d$ quarks.

\begin{figure}[thb]
\centerline{\epsfig{file=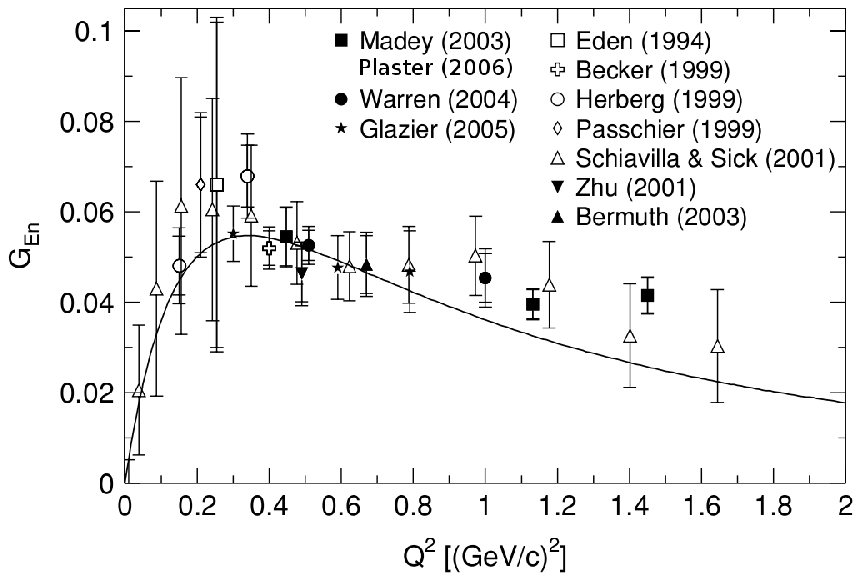,width=6.5cm,angle=0}\hspace{20mm}
\epsfig{file=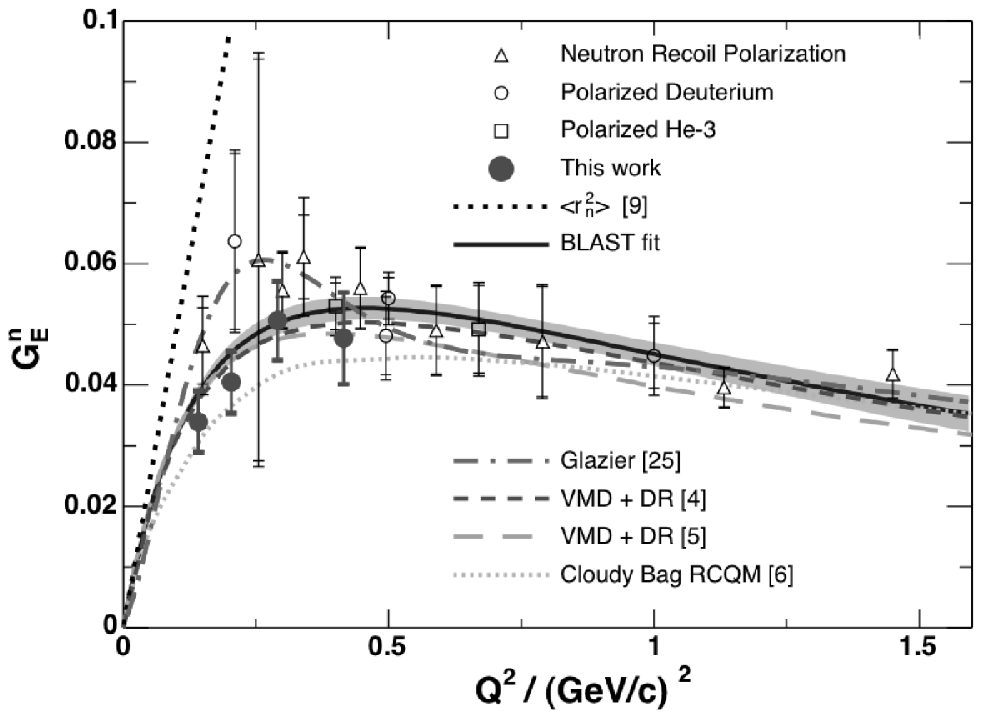,width=6.1cm,angle=0}}
  \caption{\label{fig:figure1} Left panel (from paper \protect\cite{plaster06}): The world data 
  (as of year 2006) on
$G_{En}$ versus $Q^{2}$ extracted from polarization measurements
and an analysis of the deuteron quadrupole form factor 
\protect\cite{eden94,madey03,plaster06,passchier99,glazier05,herberg99,becker99,
bermuth03,zhu01,warren04,schiavilla01}. The Galster parameterization \protect\cite{galster71} 
is shown as a solid curve.
Right panel (from paper \protect\cite{geis08}): The world data from polarization measurements (see
references in \protect\cite{geis08}). The solid circles show the results of BLAST experiment. The
new BLAST parameterization of $G_{En}$ is shown as a solid line with one-sigma error band.}
\end{figure}

A precision measurement of $G_{En}$ from quasielastic electron scattering cross
section is difficult as the neutron's net charge is zero, and $G_{En}$ is small
compared to the magnetic form factor of the neutron, $G_{Mn}$; nevertheless, the
availability of high-duty-factor polarized electron beams and polarized targets
over  last decade made possible to extract $G_{En}$ with high precision from
polarization  observables based on the interference of $G_{En}$ with $G_{Mn}$. Left
panel of Fig.~\ref{fig:figure1} shows a summary of all published data (as of year
2006) on the neutron form factors employing recoil nucleon polarimetry with
polarized electron beams and unpolarized targets
\cite{plaster06,eden94,madey03,passchier99,glazier05}, beam-target asymmetry 
measurements with polarized electron beams and polarized targets
\cite{herberg99,becker99,bermuth03,zhu01,warren04}, and an analysis combining data
on the deuteron quadrupole form factor with polarization-dependent observables
$t_{20}$ and $T_{20}$ \cite{schiavilla01}.

In the absence of a free neutron target, determinations of $G_{En}$ are typically
carried out using quasielastic electron scattering from $^2H$ or $^3He$ targets that
introduces large model-dependent  corrections and uncertainties due to
uncertainties  in  the  theoretical description  of the target nucleus, mostly from
final-state  interactions and meson-exchange currents. In addition, the
contribution from the protons in the target nucleus might introduce significant
uncertainty into the final result and should be a subject of a special attention
during experiment design and analysis periods. A flux of quasielastic protons from
deuterium or helium target is few times higher than the flux of quasielastic
neutrons; thus, relatively small misidentification of protons as neutrons  might
lead to serious proton contamination of neutron data. Though the results
dilution due to the  leak of close-to-quasifree protons through veto detectors
and analysis cuts can be reliably estimated with careful Monte-Carlo simulation of
the detector setup and verified experimentally with liquid hydrogen target, the
correction of the proton contribution from large-missing-momentum quasielastic and inelastic
reactions    is not straightforward. If the neutron detector aperture is not
surrounded with thick shielding and not protected with high magnetic field (that
deflects all protons away of the neutron detector) or large-area and high-efficiency
PID system (viz., veto that covers whole detector materials  including construction
elements), the "non-quasifree" proton might hit the neutron detector material aside
of the nominal active detector area and produce secondary particles (charged and
neutral) that will make a "fake" hit (or hits) nearby "expected-quasifree-neutron"
hit position (possibly, behind not-fired veto detector). Simulation and correction
of such effect require knowledge of exact disposition of the experimental setup
materials as well as include calculations of model-dependent polarization
observables in non-quasifree region; these corrections might introduce significant
systematic uncertainty in the final result.  


\section{Measurements at Low $Q^2$}

To study low-$Q^2$ region in a systematic manner and provide improved nucleon form
factor data,  the Bates large acceptance spectrometer toroid (BLAST) experiment was
operated at the MIT-Bates Linear Accelerator Center in 2004/2005 \cite{hasell09}.
The experiment used a longitudinally polarized ($\sim$70\%) electron beam of
$\sim$200~mA stored in the South Hall Storage  Ring; an internal polarized
($\sim$80\%) gas target of deuterium provided by an atomic beam source, and a
left-right symmetric detector based on a toroidal spectrometer with tracking
($\delta p/p \approx$ 3\%, $\delta \theta \approx$ 0.5$^o$), time-of-flight, 
Cherenkov, and neutron detectors. This advanced PID system provided  clean
identification of neutrons and protons in the detector acceptance of $\theta
\approx$ 20-80$^o$ and $\phi \approx \pm$15$^o$; and with beam-target luminosity
$L = 6 \times 10^{31}$~cm$^{-2}$s$^{-1}$, the $g \equiv G_{En}/G_{Mn}$ was extracted
from beam-target asymmetries $A^V_{ed} = P_e P_n V (a\,sin{\theta}^* cos{\phi}^* g +
b \,cos{\theta}^*)/(c g^2 + 1)$ at $Q^2$ = 0.14, 0.20, 0.29 and 0.42 (GeV/c)$^2$
with statistical uncertainty of about 12-16\% . In the formula above, $P_e$ and
$P_n$ are the beam and the target polarizations, $V$ is dilution factor,
${\theta}^*$ and ${\phi}^*$ are the target spin orientation angles with respect to
momentum transfer vector, and $a, b, and c$ are kinematic factors. A systematic
uncertainty of $\sim$6.5\% was dominated by target spin angles uncertainties.
Experiment results were published in \cite{geis08} and shown in the right panel of 
Fig.~\ref{fig:figure1}; more BLAST measurements in $Q^2$-range between 0.6 and 0.8
(GeV/c)$^2$ are expected. The distribution of all $G_{En}$ data shown in the right
panel of Fig.~\ref{fig:figure1} was parameterized by BLAST collaboration as a
function of $Q^2$ based on the sum of two dipoles, $\sum a_i / (1 + Q^2/ b_i)$ ($i$
= 1, 2). The new data from BLAST do not show a bump structure at low $Q^2$ as
suggested in \cite{glazier05} (shown as a dash-dotted line in the right panel of 
Fig.~\ref{fig:figure1}); the improved precision of the data provides strong
constraints on the theoretical understanding of the meson cloud in nucleon and
important for extraction of weak form factors.

\section{Measurements at High $Q^2$ via recoil polarimetry}

\begin{figure}[t!]
\centerline{\epsfig{file=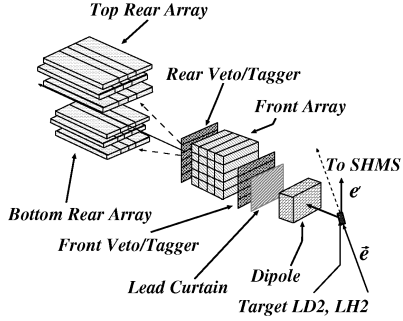,height=4.2cm,angle=0}\hspace{30mm}
\epsfig{file=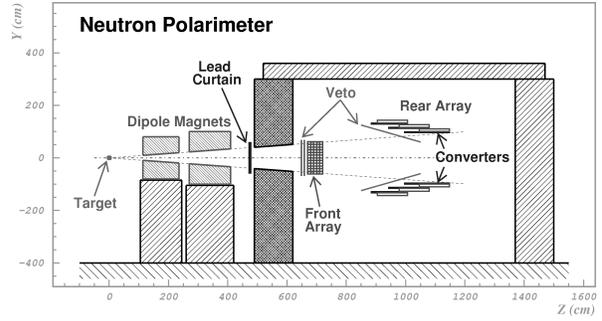,height=4.2cm,angle=0}}
  \caption{\label{fig:figure2} Left panel: Schematic diagram of the 
  experimental arrangement in E93-038.
  Right panel: Neutron polarimeter originally proposed
  for JLab E09-006 experiment.}
\end{figure}

JLab E93-038 collaboration  carried  out measurements of $G_{En}$ in 2000/2001 at 
three values of $Q^2$ (viz., 0.45, 1.13, and 1.45 (GeV/c)$^2$) \cite{plaster06}. The
reported values of the ratio of the neutron electric to magnetic form factor ratio,
$G_{En}/G_{Mn}$, represent both the highest $Q^2$ extraction and most precise published
determinations of $G_{En}$ with relative statistical uncertainties of 8.4\% and 9.5\%
at the two highest $Q^2$ points and relative systematic uncertainties of 2-3\%.
In the recoil polarimetry technique, value of $g \equiv G_{En}/G_{Mn}$ is extracted from measurements of
the neutron's recoil polarization in the quasielastic scattering of longitudinally
polarized electrons from unpolarized neutrons in deuterium; in the
one-photon-exchange approximation, $g = - (K_L/K_S) (P'_S/P'_L)$, where $P'_S$ and
$P'_L$ are the transverse and longitudinal components of the neutron's recoil 
polarization, and $K_L$ and $K_S$ are kinematic factors. Ratio of polarization
components is accessed via measurement of the ratio of up/down asymmetries of the
recoil neutron scattering in the neutron polarimeter for 2 angles of the neutron
spin precession in the dipole magnet in the front of the polarimeter (see left panel of
Fig.~\ref{fig:figure2}). A significant
advantage of the recoil polarimetry technique is that both the electron beam
polarization $P_e$ and the polarimeter analyzing power $A_y$ cancel in the
$P'_S/P'_L$ ratio, resulting in small systematic uncertainties; the nuclear
corrections are small in this ratio technique. Also, the cross-ratio technique used
for extraction of up/down asymmetries is insensitive to beam charge asymmetry and
the polarimeter geometrical asymmetry.

In 2009, JLab PAC34 approved proposal 09-006 \cite{e09-006} to extend precision
extractions of $G_{En}$ to 4.0, 5.2, and 6.9 GeV$^2$ with 12-GeV upgrade at JLab.
The experimental arrangement was similar in principle to the one used in E93-038.
With the polarized ($\sim$80\%) electron beam current of 80 $\mu$A and 40-cm liquid 
deuterium target, the expected beam-target luminosity $L = 1.02 \times 10^{39}$ is  
few orders of magnitude higher than the luminosities of competing experiments with
polarized $^3He$ target at JLab (E02-013 and E09-016).   
The  polarimeter to be used for these measurements (see right panel of
Fig.~\ref{fig:figure2})  is  an  enhanced version  of  the  one used for E93-038.
Increased vertical acceptance of the polarimeter was better matched with high-resolution electron
spectrometer (SHMS); in order to increase the polarimeter efficiency, more thick
polarimeter analyzer is used as well as 3-cm steel converters were inserted ahead of
each layer  in  the  rear detector  arrays. The thickness of the converters was
optimized to maximize the gain in the detection efficiency for neutrons. The
polarimeter was located in the bunker with the collimator that was small enough to
allow illumination of the active area of the polarimeter analyzer only; used for
spin precession 4.5-Tm dipole magnet removed almost all charged particles (including
protons) from the polarimeter acceptance. With deuteron target, missing momentum of
the quasielastic scattering can be obtained with high accuracy from high-precision
scattered electron kinematics information ($\delta p_e/p_e = 0.03-0.08$\%) and a
direction the recoil neutron trajectory without usage of not-very-accurate 
measurement of high recoil neutron momentum via TOF; an accurate cut on the missing
momentum allows efficiently select quasielastic events.    

\begin{figure}[t!]
\centerline{\epsfig{file=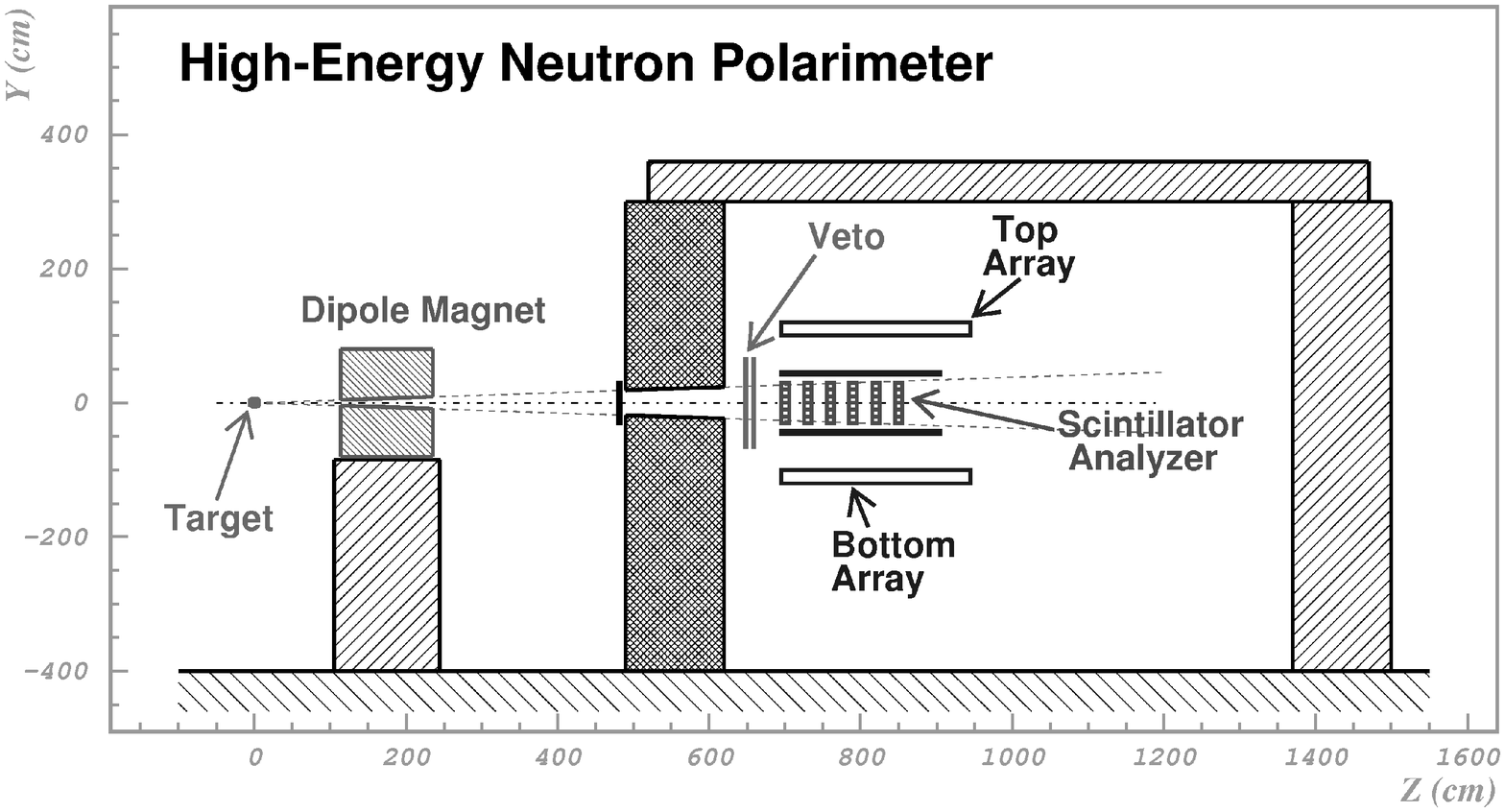,height=4.2cm,angle=0}\hspace{5mm}
\epsfig{file=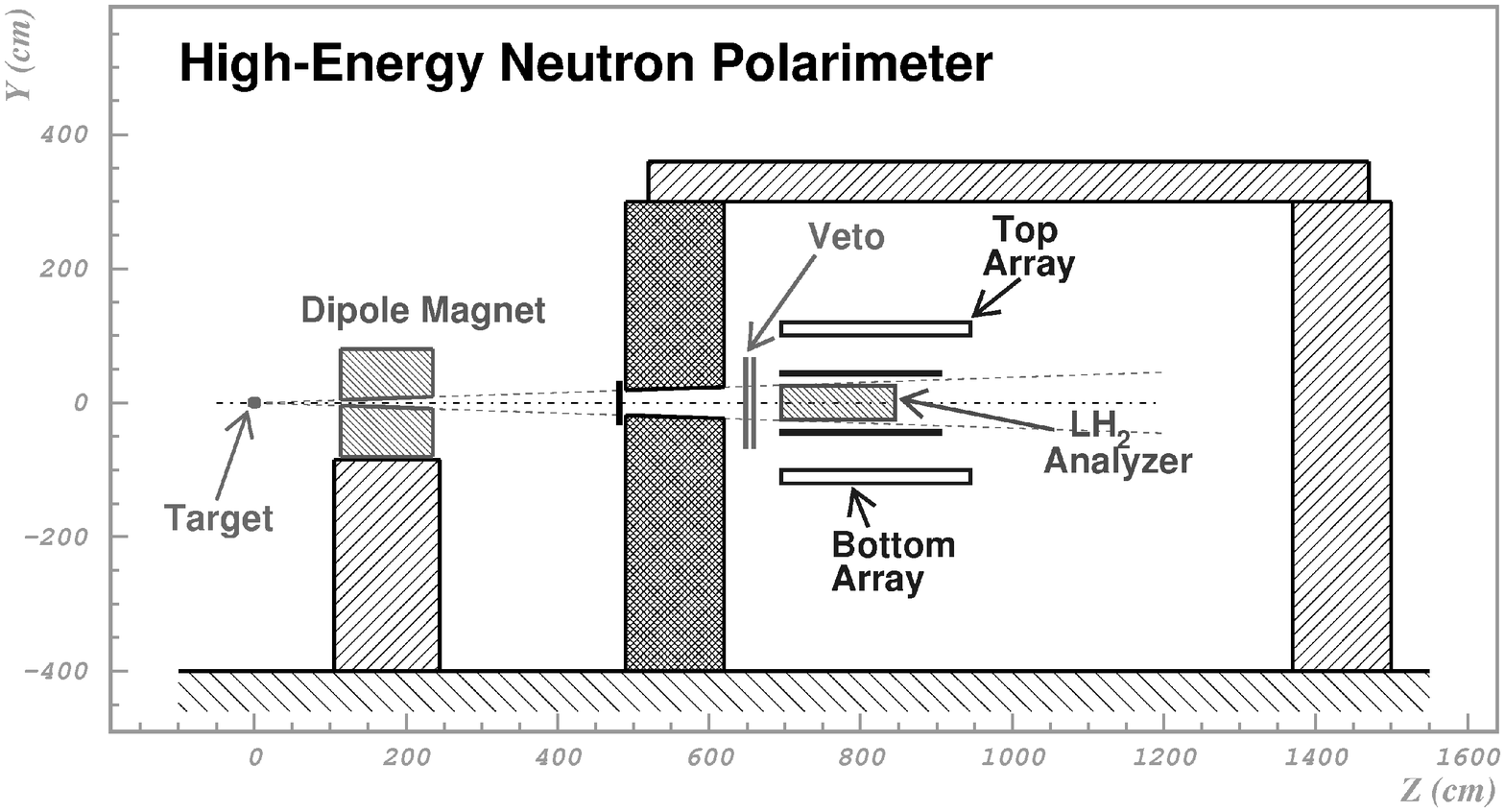,height=4.2cm,angle=0}}
  \caption{\label{fig:figure3} Left panel: Updated neutron polarimeter with
  distributed scintillator analyzer  for JLab E09-006 experiment.
  Right panel: Proposed neutron polarimeter with liquid hydrogen analyzer.}
\end{figure}

\begin{figure}[hbt]
\centerline{\epsfig{file=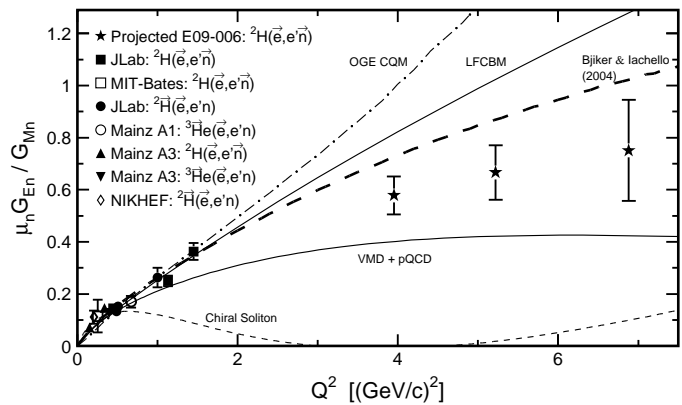,width=5.3cm,angle=270}}
  \caption{\label{fig:figure4} Predictions of selected models compared with neutron data.
  Solid-star symbols are located according to the BLAST $G_{En}$ parameterization and represent 
  projected uncertainties in 
  JLab E09-006 experiment with distributed scintillator analyzer.}
\end{figure}

Recently, to increase the efficiency of the neutron polarimeter and access neutron scattering 
at relatively small angles (where the maximum of analyzing power is located),
an updated version of the polarimeter was proposed to be used for these measurements 
(see left panel of Fig.~\ref{fig:figure3}). 
At high $Q^2$ values, the higher energy of the quasielastic neutron from   the
target allows detection of the recoil proton (instead of detection of the scattered
neutron  as was proposed in the original proposal to PAC34). In the updated
polarimeter, the  scintillation detectors of the former rear array are re-arranged
and located above and below the front array to cover the solid angle expected for
the recoil protons.  The front array consists of six columns of scintillators with a
20-cm spacing between columns.  Each column consists of one layers of six 10-cm
thick scintillators.  This segmented structure of the "scintillator analyzer"
minimizes the absorption of the recoil  proton inside the front array and provides
about 17, 21, and 26\% recoil-proton detection  efficiencies for the 4-15-degree
range of quasielastic neutron scattering at $Q^2$ = 4.0, 5.2, and  6.9 (GeV/c)$^2$;
in contrast, the neutron polarimeter in the PAC34 proposal had an efficiency  of
only 2-3\% for detecting the scattered neutron. The increased efficiency of the
polarimeter  allows to use a 60-cm high front array (instead of 120-cm-high front
array in the PAC34 proposal)  and thereby facilitates the search for the high-field
(about 4.5 Tm) dipole magnet that is needed for optimal precession of the quasielastic
neutron spin. A  double  layer of veto/tagger detectors is located ahead  of  the
front array. The thin scintillator detectors are located
in between the top/bottom arrays and the front array; analysis of the amplitudes of
signals from these detectors together with the signal amplitudes from the detectors
of the top/bottom arrays will provide $\Delta$E-E  identification of the recoil
protons (in addition to the measurements of TOF between the front and top/bottom
array detectors). High segmentation of the scintillator detectors in the polarimeter will
allow one to reconstruct reliably the polar angle of the recoil proton   with an
accuracy of 5-6 degrees that will permit controlling the angle of the neutron
scattering in the front array with an accuracy of 1.5-2 degrees (to maximize the
FOM) and practically eliminate the loss of statistics associated with  accidental
coincidences of the quasielastic neutrons and  detected background particles in the
front array. The new polarimeter increases the experiment FOM by a 
factor of about 3, and allows to reach the statistical uncertainties of about 13,
16, and 26\% at $Q^2$ = 4.0, 5.2, and  6.9 (GeV/c)$^2$ (see Fig.~\ref{fig:figure4}). The projected total 
systematic uncertainties are on the order of 3\%; a few of
the larger systematic uncertainties resulted from fluctuations in the
beam polarization (between measurements of the asymmetries at
different neutron spin precession angles) and uncertainties in the
spin precession angle.

Further dramatic improvement of the FOM of the experiment can be reached with use of
a tank with liquid hydrogen instead of the scintillator analyzer in the neutron polarimeter (see right panel
of  Fig.~\ref{fig:figure3}). A liquid-hydrogen-analyzer concept was considered more
than 10  years ago by HARP collaboration at NIKHEF to measure $G_{En}$ with high
precision  in the range of $Q^2$ in between 0.5 and 1 (GeV/c)$^2$ \cite{harp}, but that project
was not realized (because of significant cost of the project and visible absorption
of the recoil protons in the analyzer materials). In high-$Q^2$ range of E09-006, 
insignificant
absorption of the recoil protons in the liquid hydrogen compensates a relatively small
effective thickness of 150-cm-long analyzer, and well-known analyzing power for $np$
scattering is 2-2.5 times higher than one for $CH_2$ \cite{ladygin99} that results
in the statistical uncertainties of about 6 and 15\% at $Q^2$ = 4 and 6.9
(GeV/c)$^2$. Most probably, long and expensive development and production will be
needed for about 300-litter $LH_2$ tank to satisfy the safety requirements in
Jefferson Lab; nevertheless, advantages of use of the liquid-hydrogen analyzer
remove all principal limits of $G_{En}$ measurements at even higher $Q^2$ values.   

\section{Measurements at High $Q^2$ via beam-target asymmetry}
In 2006, E02-013 collaboration at JLab conducted measurements of $G_{En}$ in $Q^2$
range between 1.2 and 3.4 (GeV/c)$^2$ via measurements of asymmetries of recoil
neutron fluxes from interaction of polarized ($\sim$75\%) electron beam with 
polarized ($\sim$50\%) $^3He$ target. Main idea of the experiment was to match large
acceptance of scattered electrons detection in the BigBite spectrometer
($\sim$76~msr) with large acceptance of the neutron detector ($\sim$80~msr).
Relatively small (about 1250 times lower than in the JLab E09-006 experiment) beam-target luminosity  
of about $8 \times 10^{35}$~cm$^{-2}$s$^{-1}$ (that corresponds to a 12 $\mu$A electron beam with 40-cm $^3He$
gas target at about 10 atmospheres pressure) was declared in the original proposal \cite{e02-013},
and was limited by abilities of charge track reconstruction system of open-acceptance BigBite spectrometer.
With this declared luminosity, the expected statistical uncertainty was about 14\% at
high $Q^2$ point. 
The final results of this experiment are not published yet;
nevertheless, neutron detector in E02-013 experiment had no neutron aperture collimator and was not 
shielded from charged particles with magnet field; open exposure of the neutron detector to the 
intensive flux of protons from the target
might lead authors to reconsider the expected in the original proposal
systematic uncertainty of about 10\%. Direct comparison of the $G_{En}$ result from
E02-013 at $Q^2$ in the range of 1.2-1.5 (GeV/c)$^2$ with the high-precision
measurements of E93-038 experiment as well as with the result of recently conducted
experiment of A1 group at Mainz with polarized $^3He$ target at $Q^2$ = 1.5 (GeV/c)$^2$ \cite{mami05} might 
help estimate systematics in E02-013 measurements.    

In 2009, JLab PAC34 approved proposal 09-016 \cite{e09-016} to extend measurements 
of $G_{En}$ with $^3He$ target to 5.0, 6.8, and 10.2 GeV$^2$ with 12-GeV beam upgrade 
at JLab, upgraded Super-BigBite spectrometer, and enhanced neutron detector located at the small angle of 17
degrees in respect to the beam direction. With significantly higher
than in E02-013 electron beam current of $\sim$65~$\mu$A, authors expect to reach the 
statistical  uncertainty of about 20\%; the expected systematic uncertainty is about 8\%. 

\section{Conclusions}
Recently conducted and planning $G_{En}$ experiments at Jefferson Lab, MIT and Mainz 
will extend the covered $Q^2$ range to be similar to the $Q^2$ range covered by
other electromagnetic nucleon form factor measurements. Direct comparison of the
experimental $G_{En}$ results is required for analysis of systematic uncertainties
introduced by different measurement techniques. 

\section{Acknowledgments}
The author wish to thank the members of Jefferson Lab E09-006 collaboration for the
support and helpful discussions during the experiment planning.

\end{document}